# Dipolar excitonic insulator in a moiré lattice

Jie Gu[1], Liguo Ma[1], Song Liu[2], Kenji Watanabe[3], Takashi Taniguchi[3], James C. Hone[2], Jie Shan[1,4,5*], Kin Fai Mak[1,4,5*]

[1]School of Applied and Engineering Physics, Cornell University, Ithaca, NY, USA
[2]Department of Mechanical Engineering, Columbia University, New York, NY, USA
[3]National Institute for Materials Science, 1-1 Namiki, 305-0044 Tsukuba, Japan
[4]Laboratory of Atomic and Solid State Physics, Cornell University, Ithaca, NY, USA
[5]Kavli Institute at Cornell for Nanoscale Science, Ithaca, NY, USA

*Email: jie.shan@cornell.edu; kinfai.mak@cornell.edu

**Two-dimensional (2D) moiré materials provide a highly-controllable solid-state platform for studies of correlated quantum phenomena [1-4]. To date, experimental studies have focused on the correlated electronic states [5-13]; the correlated bosonic states in moiré materials have remained practically unexplored. Here, we report a correlated dipolar excitonic insulator--a charge insulating state driven by exciton formation [14, 15]--in a Coulomb-coupled $WSe_2$ monolayer and $WSe_2/WS_2$ moiré bilayer at total hole doping density equal to the moiré density. The system is a Mott insulator when all the holes reside in the moiré layer. Under an out-of-plane electric field, the holes can be continuously transferred to the $WSe_2$ monolayer, but remain strongly bound to the empty moiré sites. This is effectively an interlayer exciton fluid in the moiré lattice under a particle-hole transformation. We identify the phase space and determine the charge gap energy of the excitonic insulating state by optical spectroscopy and capacitance measurements, respectively. We further observe the emergence of local magnetic moments in the $WSe_2$ monolayer induced by the strong interlayer Coulomb correlation. Our result paves the path for realizing correlated bosonic quantum phenomena described by the Bose-Hubbard model in a solid-state system [16-20].**

We consider a heterostructure of a $WSe_2$ monolayer and a $WSe_2/WS_2$ moiré (bi)layer that are separated by a thin hexagonal boron nitride (hBN) spacer (Fig. 1a). The spacer suppresses single-particle tunneling [21, 22] between the monolayer and the moiré layer, but is sufficiently thin (2-4 layers, thickness $d \sim 1$ nm) relative to the moiré period ($a_m \sim 8$ nm) so that the interlayer Coulomb interaction is as important as the Coulomb interaction within the moiré layer. We discuss the case of hole doping; the concept also applies to the case of electron doping. The valence band edge of the $WSe_2/WS_2$ bilayer resides in the constituent $WSe_2$ layer [10, 11]; it is above the band edge of the $WSe_2$ monolayer because of the moiré potential [12] (Fig. 1b). The flat moiré bands in angle-aligned $WSe_2/WS_2$ bilayers have been shown to realize the triangular lattice Fermi-Hubbard model in the strong correlation regime [9-12, 23, 24]. When the moiré layer is populated with one hole per moiré site and the monolayer is undoped, the system is an electronic Mott insulator (Fig. 1d, left). An out-of-plane electric field, $E$, which shifts the relative band alignment, can be tuned to transfer some holes from the moiré layer to the monolayer without changing the total hole density (Fig. 1d, middle). The holes in the monolayer are bound to the empty moiré sites to minimize the Coulomb repulsion energy. Under a particle-hole transformation in the moiré layer, the empty sites become electrons. The resultant bound electron-hole pairs, which are strongly interacting, can hop around and form an interlayer exciton fluid in the moiré lattice (Fig. 1d, right). It is a zero-magnetic-field analogue of the exciton fluid in quantum Hall bilayers [25-27], induced by the moiré flat bands instead of the dispersionless Landau levels.

The interlayer excitons are tightly bound in the limit of low exciton density and $d \ll a_m$. Ignoring extended Coulomb interactions, dissociation of an exciton corresponds to removing a hole from its pairing empty site and creating a doublon on another site. We estimate the exciton binding energy using the onsite Coulomb repulsion energy, $\sim \frac{e^2}{4\pi\epsilon\epsilon_0 a_m} \sim 50$ meV (here $e$, $\epsilon \approx 4.5$ [12] and $\epsilon_0$ denote the elementary charge, hBN effective dielectric constant and vacuum permittivity, respectively; and $a_m$ approximates the hole Wannier function size). Driven by the exciton formation, the ground state is charge-incompressible [14, 15] but exciton-compressible in general [21]; the exciton density can be continuously varied by the electric field $E$. The new charge insulating state is distinct from the Mott insulating state in an isolated moiré layer; the empty sites introduced by the Coulomb-coupled monolayer enables exciton hopping. The exciton fluid is expected to be a 2D quasi-condensate and superfluid at sufficiently low temperatures and disorder densities [28-30]. The large exciton binding energy favors high transition temperatures [31, 32]. Compared to exciton fluids in a continuum [21], the moiré flat bands enhance the exciton-exciton correlation and further stabilize these states.

In this study, we demonstrate an experimental realization of equilibrium excitons in a lattice using the platform of moiré heterostructures. Figure 1a illustrates the experimental device structure. Both 0- and 60-degree-aligned WSe$_2$/WS$_2$ bilayers (moiré density $\approx 1.9 \times 10^{12}$ cm$^{-2}$) have been studied; they show similar results. The heterostructure is dual-gated with nearly symmetric top and bottom gates made of hBN gate dielectrics and graphite gate electrodes. The two gate voltages ($V_t$ and $V_b$ for the top and bottom gates, respectively) tune independently the total hole density $\nu$ (measured in units of the moiré density) through ($V_t + V_b$) and the out-of-plane electric field $E$ through ($V_t - V_b$). The holes are generally distributed in both the moiré layer ($\nu_m$) and the 'moiré-free' monolayer ($\nu_f$): $\nu = \nu_m + \nu_f$. We probe the correlation effects as a function of ($\nu, E$) optical spectroscopies and determine the charge gap of the correlated insulating states by capacitance measurements. We further demonstrate charge localization and emergence of local magnetic moments in the monolayer induced by the interlayer Coulomb correlation through magnetic circular dichroism (MCD) measurements. Details on the device fabrication and measurement methods are provided in Methods. Unless otherwise specified, all measurements are performed at 3.4 K.

Figure 1e, f illustrate the doping dependence of the optical reflection contrast and photoluminescence (PL) spectra of the heterostructure, respectively, at $E = -34$ mV/nm. We can identify in the reflection spectrum resonances from moiré excitons [33-39] and the 1s and 2s excitons of the monolayer (Fig. 1c is the spectrum at zero-doping). The PL around

1.4 eV arises from recombination of holes in $WSe_2$ and electrons in $WS_2$ of the moiré layer [11, 40] (linecuts at different $\nu$'s are shown in Extended Data Fig. 2). The PL intensity generally decreases with doping. At $E$ = -34 mV/nm, holes are doped only into the moiré layer and the monolayer remains charge neutral for the entire doping range in Fig. 1e, f. This is supported by the $\nu$-indepndent 1s exciton response of the monolayer. The 2s exciton of the monolayer has been demonstrated to be a sensitive probe of the charge insulating states in proximate materials [12]; it manifests a spectral blueshift with enhanced reflection contrast due to reduced dielectric screening of the excitonic interactions. We clearly identify five insulating states in the moiré layer, including the band insulator ($\nu$ = 0), Mott insulator ($\nu$ = 1), and electron crystals ($\nu$ = 1/3, 2/3, 1/2) in order of decreasing charge gap size. But the 2s exciton loses transition amplitude and sensitivity to correlated insulating states upon doping in the monolayer [12].

The correlated insulating states can also be identified by PL of the moiré layer with suppressed emission intensity and abrupt energy shift [11, 40]. The latter is presumably related to the chemical potential jump with doping across a correlated insulating state. Further theoretical studies are required to fully understand the microscopic mechanism. The spectral responses below $\nu \sim 1/3$ are likely distorted by the nonlinear gating effect from the non-ideal electrical contact to the heterostructure at low doping densities and low temperatures. Below we explore the ($\nu, E$) parameter space for new correlated states using PL as a probe.

Figure 2a, b show the doping dependence of the reflection spectrum of the monolayer 1s exciton and the PL spectrum of the moiré layer, respectively, at four representative electric fields. At $E$ = -10 mV/nm, the spectra for $\nu \leq 1$ resemble that of Fig. 1e, f. For $\nu$ > 1, the 1s exciton turns into an attractive polaron (redshifted branch, denoted as $X^+$) and a repulsive polaron (blueshifted branch, not labeled), which signal hole doping into the monolayer [22, 41]. The reflection change at $\nu$ = 1 is correlated with a local intensity minimum and an abrupt energy shift of the PL. As $\nu$ further increases, the PL follows a smooth doping dependence till $\nu \approx 1.8$ while $X^+$ remains nearly unchanged. These results indicate that the holes are first doped into the moiré layer to completely fill the first Hubbard band; further doping is introduced into the monolayer till $\nu \approx 1.8$, at which the chemical potential reaches the second Hubbard band of the moiré layer. The state at $\nu$ = 1 is an electronic Mott insulator. For $\nu$ between 1 and 1.8, the doping density in the moiré layer is fixed at $\nu_m = 1$.

At larger electric fields ($E$ = 15 and 18 mV/nm), the polaron feature ($X^+$) appears at smaller $\nu$'s (< 1) in the reflection spectrum; accordingly, the extended region of $\nu_m = 1$ emerges at larger $\nu$'s (> 1) in the PL spectrum. In addition, the electron crystal states can no longer be identified. These results support that the holes are now shared between the monolayer and the moiré layer before the density in the moiré layer reaches $\nu_m = 1$. The electron crystal states are quantum melted [23] because the extended Coulomb repulsion in the moiré layer that induces these states are screened by the holes in the monolayer.

Remarkably, we observe an abrupt energy shift in both the attractive polaron resonance and the PL at total filling $\nu = 1$. This signifies a chemical potential jump [11, 42] and emergence of a new correlated insulating state. Such a state is unexpected if the monolayer and the moiré layer are uncoupled. In that case both layers should be charge-compressible. The state is consistent with the dipolar excitonic insulator induced by the strong interlayer Coulomb interaction (Fig. 1d): holes in the monolayer ($\nu_f$) and in the moiré layer ($\nu_m$) with $\nu_m + \nu_f = 1$ form effective dipolar excitons with exciton density $\nu_X = \nu_f$. The exciton density increases with electric field; the exciton binding energy decreases and the excitonic insulator weaken. The state can no longer be observed (e.g. at $E \gtrsim 30$ mV/nm) beyond the exciton Mott density, at which the excitons turn into an electron-hole plasma [21, 31].

The $(\nu, E)$ parameter space is mapped out for the excitonic insulating state (Fig. 3a, b). It can be divided into four electrostatic doping regions: (I) the monolayer is charge neutral ($\nu_f = 0$); (II) the monolayer is doped and the moiré layer is at $\nu_m = 1$; (III) doping density in both layers is $< 1$ (region of interest); (IV) the moiré layer is charge neutral ($\nu_m = 0$). We use two spectral responses to identify these regions. The amplitude of the 1s exciton reflection contrast is sensitive to doping in the monolayer [22, 41]. We determine the boundary between the neutral and doped monolayer (dash-dotted lines) in Fig. 3b. The PL exhibits an intensity minimum when the Fermi level touches a Hubbard band (both top and bottom) of the moiré layer. We determine the boundaries of the first Hubbard band (dashed lines) in Fig. 3a. Details on the doping region assignment and maps of other spectral responses, including the PL peak energy and the attractive polaron reflection contrast, are provided in Methods and Extended Data Fig. 6, 7.

The optical studies (Fig. 3a) suggest an excitonic insulator over a large range of electric field and a narrow range of doping density around total filling $\nu = 1$ in Region III. We characterize the state quantitatively by capacitance measurements. The capacitance device is similar to that of Fig. 1a with a local bottom gate (Extended Data Fig. 4a), which allows electrostatic doping of the uncovered heterostructure area (the contact area) by the top gate to achieve good electrical contacts [43]. The accessible range of $(\nu, E)$ is thus limited in order to maintain good electrical contacts. The differential gate capacitance, $C$, is measured by collecting charges on the bottom gate while applying a small AC voltage on the heterostructure. We expect $C \approx C_g$ if the heterostructure is charge-compressible (conducting), and $C < C_g$ if it is charge-incompressible (insulating), with $C_g$ denoting the geometrical bottom-gate-to-sample capacitance. Figure 3c is the measured $C/C_g$ as a function of $(\nu, E)$ at 20 K (good electrical contacts are achieved at the measurement frequency of 443 Hz). The boundaries for different electrostatic doping regions (dashed lines) are obtained from optical measurements on the same device. In Region III, we observe $C < C_g$ around total filling $\nu = 1$ over an extended range of electric field, and $C \approx C_g$ in the background. The charge-incompressible state revealed by capacitance is fully consistent with that from the optical studies of Fig. 3a and Extended Data Fig. 7. (The larger electric fields here compared to Fig. 3a, b are due to the thinner hBN spacer in the capacitance device.) The state persists up to ~ 120 K. Details on the capacitance measurements and results at other temperatures are included in Methods and Extended Data Fig. 4, 5.

We evaluate the charge gap of the excitonic insulator by integrating the capacitance dip [43], $\frac{e}{2} \int dV_b \left(1 - \frac{c}{c_g}\right)$. Figure 3d summarizes the electric-field dependence of the charge gap at varying temperatures (in contrast, the charge gap of the electronic Mott insulator in bare WSe$_2$/WS$_2$ moiré bilayers is nearly independent of $E$ [11]). The top axis shows the corresponding exciton density, $\nu_X$, from electrostatics calculations. The smooth electric-field dependences show that the exciton fluid is compressible (that is, the holes can be continuously transferred between the layers). The gap energy at small exciton density and low temperature ($\sim 30$ meV) provides a measure of the exciton binding energy [21], which is consistent with our estimate based on the on-site repulsion energy. The charge gap vanishes at large $\nu_X$'s; at low temperature, it is around 0.7 (or $\sim 1.4 \times 10^{12}$ cm$^{-2}$, also see Extended Data Fig. 7). This provides an estimate of the Mott density, beyond which the charge-incompressible excitonic insulator transitions into a charge-compressible electron-hole plasma at low temperatures [21, 31]. The large exciton binding energy, high Mott density and ionization temperature observed in our experiment are all consistent.

Finally, we demonstrate charge localization and emergence of local magnetic moments in the monolayer induced by the strong interlayer Coulomb correlation in the excitonic insulating state. We measure the magnetic response of the monolayer under an out-of-plane magnetic field by MCD near its repulsive polaron feature, which has been shown to be proportional to the sample magnetization [11, 44, 45]. Figure 4a shows the magnetic-field dependence of the MCD at $\nu_X \approx 0.3$. The behavior at other exciton densities is similar. At 3.4 K, the MCD increases linearly with magnetic field under small fields, and saturates above $\sim 2$ T. The saturation field increases rapidly with increasing temperature. The temperature dependence of the magnetic susceptibility, $\chi$, extracted from the zero-field slope of the MCD [11, 44], follows the Curie-Weiss behavior $\chi^{-1} \propto T - \theta$ (line in Fig. 4b). Here $T$ and $\theta$ denote the sample temperature and the Weiss constant, respectively. Such a behavior is in contrast to that of an isolated monolayer at comparable hole doping densities [11, 45, 46] but compatible with the emergence of local magnetic moments [11, 44]. The Weiss constant obtained from the fits (solid lines) changes from $\sim 0$ to -2 K when the exciton density changes from $\nu_X \approx 0.3$ to 0.45. It suggests a weak antiferromagnetic exchange interaction between the distant local moments in the monolayer.

In conclusion, we have realized equilibrium exciton fluids in a lattice using the platform of semiconductor moiré heterostructures. The exciton fluid emerges in a Coulomb-coupled WSe$_2$ monolayer and a WSe$_2$/WS$_2$ moiré bilayer at combined doping density equal to the moiré density; it is stabilized by the strong interlayer Coulomb correlation induced by the moiré flat bands. Complementary optical spectroscopy and capacitance studies have established the charge insulating character of the exciton fluid; future studies are required to establish the expected superfluid ground state. Our results pave the path for the realization of exotic correlated bosonic states, such as excitonic Mott insulator [16, 17], Wigner solid [18] and supersolid [18-20].

**Methods**

**Device fabrications**

We fabricate dual-gated devices of transition metal dichalcogenide (TMD) heterostructures using the reported layer-by-layer dry transfer technique [47]. Details have been reported in earlier studies [11, 12, 48]. The capacitance devices are fabricated following the design in Ref. [43]. Figure 1a shows the cross-section schematic of the devices. Extended Data Fig. 1 shows an optical image of device 1, the optical studies of which are presented in the main text. In particular, the $WSe_2$ monolayers are exfoliated from bulk crystals grown by the flux growth method [49]; and the $WS_2$ monolayers are from bulk crystals grown by the chemical vapor transport method (from HQ Graphene). Angle alignment of the $WSe_2$ and $WS_2$ monolayers in the moiré bilayer is achieved with assistance of the second harmonic generation [11]. Both 0- and 60-degree-aligned $WSe_2/WS_2$ bilayers are involved in this study; no noticeable difference is observed in the two cases. The 'moiré-free' $WSe_2$ monolayer is not angle aligned. The thickness of the hBN spacer between the moiré layer and the $WSe_2$ monolayer varies between 2 and 4 layers in different devices. No obvious dependence on the spacer thickness is observed, which is consistent with significant suppression of interlayer tunneling as well as negligible moiré printing on the monolayer. The top and bottom gates are nearly symmetric with hBN gate dielectrics and few-layer graphite gate electrodes; the thickness of the gate dielectrics is typically between 10 and 40 nm. Few-layer graphite is also used as contact electrodes in devices used for the optical studies. Pt electrodes are used to achieve better p-type contacts to the TMD heterostructures for the AC capacitance studies.

**Optical measurements**

The devices are characterized by the optical reflection, magnetic circular dichroism (MCD) and photoluminescence (PL) spectroscopies. They are mounted in an AttoDry 1000 cryostat. In the optical reflection measurements [11, 12], we focus white light from a tungsten halogen lamp onto the devices with a micron spot size using an optical microscope objective with numerical aperture 0.8. The incident power is less than 1 nW. The reflected light from the devices is collected by the same objective and sent to a spectrometer equipped with a charge-coupled device (CCD) for spectrum analysis. To obtain the reflection contrast spectrum, we use as a reference the reflected light intensity from the sample at a large hole or electron doping density, for which all the exciton features become negligible. We use the same setup for the MCD measurements in the polar geometry; the magnetic field is out-of-plane; the incident white light is circularly polarized (details see [11, 44]). The MCD spectrum is defined as the difference between the left- and right-handed reflected light intensity divided by the total reflected light intensity. Extended Data Fig. 3 shows the MCD spectrum as a function of magnetic field at several representative electric fields and fixed total doping density $\nu = 1$. We focus on the spectral range of the attractive and repulsive polarons of the monolayer. No MCD signal is observed at zero magnetic field; the MCD increases monotonically with magnetic field and saturates around 2-3 T. The MCD is resonantly enhanced near the energy of both the repulsive and attractive polarons; it changes sign across the polaron resonances, which is consistent with the reported results [11, 44]. We integrate the MCD signal over a narrow spectral window (boxed) near the repulsive polaron resonance to obtain Fig. 4 in the main text. The integrated MCD is proportional the valley magnetization of the carriers in the WSe$_2$ monolayer [11, 44]. For the PL measurements, we employ a He-Ne laser (633 nm) as the excitation source; the incident power is 30 nW.

**Capacitance measurements**

We follow the design reported in Ref. [43] to achieve good electrical contacts to the TMD heterostructure for the capacitance measurements. Extended Data Fig. 4a shows the schematic. The top gate (the global gate) covers the entire TMD heterostructure; the bottom gate (the local gate) covers it partially. The overlapped area is the region of interest; the non-overlapped area is the contact region. A negative voltage is applied to the global gate to introduce holes into the entire TMD heterostructure; the bottom local gate depletes holes in the region of interest (red box). The device structure limits the range of $(\nu, E)$ that can be accessed in the capacitance study because a finite negative top gate voltage is required to maintain hole doping in the contact region to achieve good electrical contacts.

The TMD heterostructure is DC-grounded. An AC excitation voltage (amplitude = 10 mV and frequency = 443-3317 Hz) is applied to the TMD heterostructure. Charge is collected from the bottom gate and sent to an on-chip high-electron-mobility transistor (HEMT) amplifier and lockin for detection [50]. The differential capacitance $C$ is the ratio of the collected charge to the excitation voltage. We verify that the result is largely independent of the excitation frequency.

Extended Data Fig. 4b shows $C$ in the units of $C_g$ as a function of $(\nu, E)$ at 20 K, where $C_g$ is the geometrical capacitance defined by the thickness and dielectric constant of the hBN layer between the bottom gate and the TMD heterostructure. The electrostatic doping regions are determined by optical responses in Extended Data Fig. 4c,d. Only region III and IV can be accessed. The diagonal dashed green line corresponds to constant top-gate voltage -1.4 V. It marks the onset of charging into the device. Top-gate voltage above the green line is required to dope the contact region and turn on the channel for capacitance measurements.

Extended Data Fig. 4c shows the PL peak intensity from the moiré layer as a function of $(\nu, E)$. The PL intensity is sensitive to doping in the moiré layer [11, 40]. The dashed line marks the boundary between Region III and IV, corresponding to doped and neutral moiré layer, respectively. Extended Data Fig. 4d shows the peak reflection contrast of the monolayer 1s exciton, which is sensitive to doping in the monolayer [22, 41]. The black dashed line marks the onset of hole doping in the monolayer. It is consistent with the onset of finite capacitance signal (Extended Data Fig. 4b). A change in the 1s reflection contrast is also observed across the green dashed line, which marks the onset of charging into the device. Detailed discussions on the electrostatic doping regions are provided below.

The measured differential capacitance accesses the quantum capacitance, $C_Q$, of the TMD heterostructure, which represents its electronic compressibility, by $\frac{1}{C} \approx \frac{2}{C_Q} + \frac{1}{C_g}$ [43]. We expect $C \approx C_g$ for a charge-compressible state and $C < C_g$ for a charge-incompressible state. A charge-incompressible state is clearly observed at $\nu = 1$ over an extended range of electric field. The state weakens with increasing electric field. This is fully consistent with the optical results (Extended Data Fig. 7). The absence of a charge-incompressible state at the moiré filling $\nu_m = 1$ also shows negligible moiré printing onto the monolayer from the moiré layer.

We perform temperature dependence studies of the capacitance. Extended Data Fig. 5 shows the results at 20, 40, 60, 80, 120, and 150 K. With increasing temperature, the charge-incompressible state at $\nu = 1$ becomes weaker and vanishes around 120 K. This corresponds to thermal ionization of the excitons; the exciton fluid turns into electron-hole plasma, which are charge-compressible [21]. The charge-incompressible state at smaller electric field (corresponding to smaller exciton density) persists to higher temperatures, that is, the exciton ionization temperature or binding energy increases with decreasing density. This is consistent with the exciton Mott transition, at which the exciton binding energy drops to zero [21, 31, 51].

**Electrostatics phase diagram**

Extended Data Fig. 6 shows the electrostatics phase diagram with a larger parameter range for $(\nu, E)$ than Fig. 3 in the main text. Extended Data Fig. 6a-c are the (peak) PL intensity of the moiré layer, the reflection contrast of the 1s exciton and the attractive polaron of the WSe$_2$ monolayer, respectively. The dashed lines outline the boundaries of different electrostatic doping regions. Below we describe how these boundaries are determined from the spectral responses.

We first examine the 1s exciton reflection (Extended Data Fig. 6b). The feature is sensitive to doping in the monolayer [22, 41]. We can draw the boundary between the charge-neutral (red) and hole-doped region (blue) based on the 1s response. The boundary is also schematically shown in Extended Data Fig. 6d as the blue line. Along the line the monolayer remains charge neutral ($\nu_f = 0$) and holes are doped solely into the moiré layer ($\nu = \nu_m$). The line traces the valence band edge of the monolayer as it shifts across the Hubbard bands of the moiré layer under an applied electric field. It therefore probes the chemical potential as a function of doping density of the moiré layer. In particular, the boundary is vertical over a range of electric field, $\Delta E$, when the monolayer band edge is located inside a charge gap of the moiré layer. The product of $\Delta E$ and the separation between the monolayer and the moiré layer, $d$, corresponds approximately to the charge gap energy. When the monolayer band edge overlaps with a Hubbard band, the boundary line gains a finite slope, which is inversely proportional to the thermodynamic density of states of the particular Hubbard band. Our data shows that the $\nu_m = 1$ Mott gap is about twice the $\nu_m = 2$ moiré band gap, and the first Hubbard band is flatter than the second. These results are fully consistent with the reported properties of angle-aligned WSe$_2$/WS$_2$ bilayers [12].

The map of the attractive polaron response is expected to be complementary to that of the 1s exciton since a charge-neutral monolayer gives rise to strong 1s exciton and weak attractive polaron reflection [22, 41]; these spectral responses are reversed for a doped monolayer. This is consistent with our experimental results of Extended Data Fig. 6b,c.

Next we examine the PL intensity from recombination of holes in $WSe_2$ and electrons in $WS_2$ of the moiré layer (Extended Data Fig. 6a). Large PL intensity (red) is observed when the moiré layer is charge neutral ($\nu_m = 0$). A local PL intensity minimum emerges when the Fermi level touches a Hubbard band. The results are consistent with recent studies [11, 40]. We draw the boundaries of the Hubbard bands (both their top and bottom) by following the PL local minimum. The boundaries of the first and second Hubbard bands are also shown schematically as orange lines in Extended Data Fig. 6d. The solid and dashed lines denote the top and bottom of a Hubbard band, respectively. Similar to the analysis of the 1s exciton above, the boundaries correspond to constant filling of the moiré layer at $\nu_m = $ 0, 1, or 2 and doping into the monolayer with $\nu_f = \nu - \nu_m$. Each orange line traces a Hubbard band edge as it moves relative to the monolayer valence band. It probes the thermodynamic equation of state of the monolayer. Because there is only one semiconductor charge gap of ~ 2 eV in the monolayer, each Hubbard band edge gives only one vertical line when it moves inside the monolayer gap. Outside the monolayer band gap, the finite slope of the orange line is inversely proportional to the thermodynamic density of states of the monolayer.

We summarize the distinct regions of the electrostatics phase diagram shown in Extended Data Fig. 6.

- Region I: $\nu_f = 0$ and $\nu = \nu_m$. No electric field dependence is seen here because all of the holes reside in the moiré layer. Electron crystal states can be clearly identified as vertical straight lines at fractional filling factors. The PL features below $\nu \sim 1/3$ are likely artifacts from the nonlinear charging effect.
- Region II: $\nu_m = 1$ and $\nu_f = \nu - 1$. Here the Fermi level is inside the Mott gap of the moiré layer.
- Region III (region of interest): $\nu_f > 0$ and $\nu_m < 1$. Both layers are hole-doped and the applied electric field transfers holes between the layers. The electron crystal states disappear in this region because of hole screening from the $WSe_2$ monolayer.
- Region IV: $\nu_m = 0$ and $\nu = \nu_f$.

Similarly, we can assign regions I', II' and III', which are the second Hubbard band copies of regions I, II and III. They correspond to ($\nu_f = 0$ and $\nu = \nu_m$), ($\nu_m = 2$ and $\nu_f = \nu - 2$) and ($\nu_f > 0$, $1 < \nu_m < 2$), respectively.

**Figures**

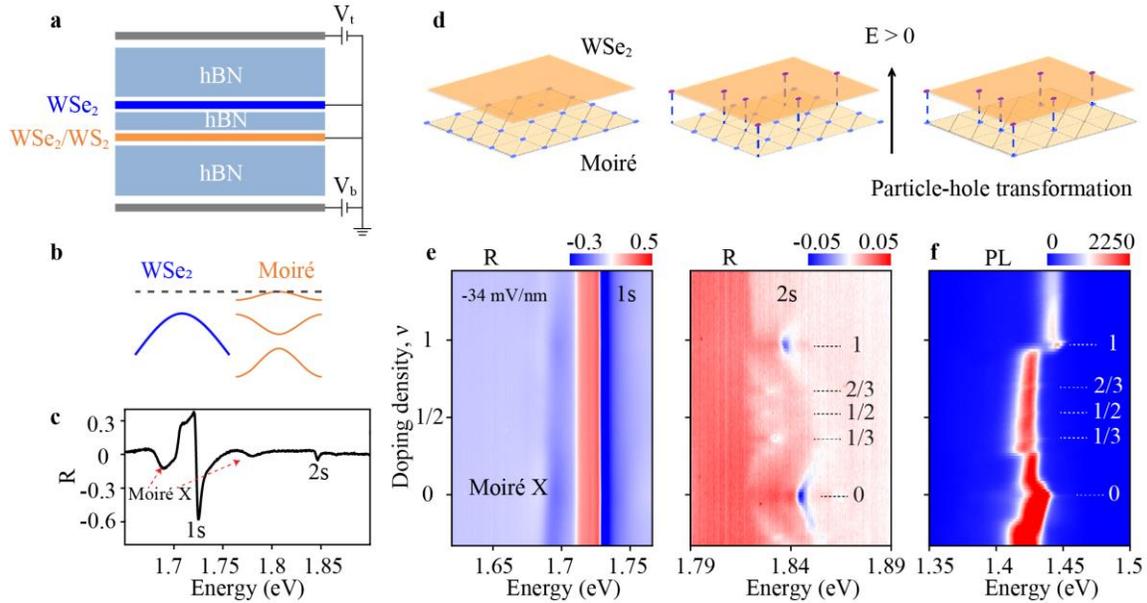

**Figure 1 | Semiconductor heterostructures for realization of a dipolar exciton fluid in a lattice. a.** Schematic of a dual-gated heterostructure of a WSe$_2$ monolayer and a WSe$_2$/WS$_2$ moiré layer that are separated by 2-4 layers of hBN. The top and bottom gate voltages ($V_t$ and $V_b$) control the doping density, $\nu$, and the out-of-plane electric field, $E$, in the heterostructure. **b.** Schematic valence band alignment of the heterostructure in the absence of electric field. The dashed line denotes the Fermi level. **c.** Reflection contrast spectrum at $\nu = 0$ and $E = $ -34 mV/nm from device 1, showing two moiré excitons and the 1s and 2s excitons of the monolayer. **d.** Left: electronic Mott insulator at $\nu = 1$ (one hole per moiré site) when all the holes reside in the moiré layer in the absence of electric field; Middle: some of the holes are transferred to the WSe$_2$ monolayer under a positive electric field. They are tightly bound to the empty moiré sites to minimize the total Coulomb repulsion energy. Right: the empty sites in the moiré layer are equivalent to electrons under a particle-hole transformation. A dipolar exciton fluid emerges in the moiré lattice. **e, f.** Reflection contrast (**e**) and PL (**f**) spectrum of the heterostructure as a function of doping density at $E = $ - 34 mV/nm. The charge insulating states are identified by both reflection contrast and PL.

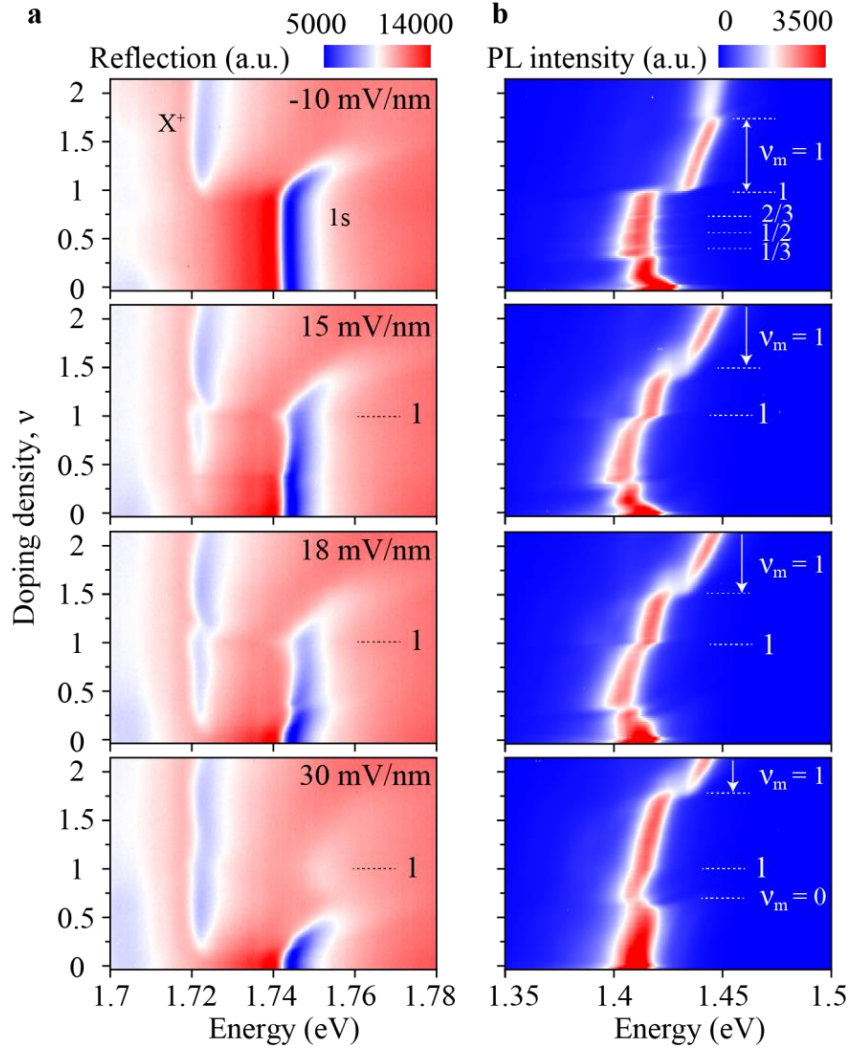

**Figure 2 | Optical responses under an out-of-plane electric field.** Optical reflection spectrum of the monolayer 1s exciton (**a**) and PL spectrum of the moiré layer (**b**) as a function of doping density from device 1. The out-of-plane electric field is -10, 15, 18, and 30 mV/nm, respectively, from the top to the bottom row. Upon doping in the monolayer, the 1s exciton feature turns into an attractive polaron (redshifted branch, X$^+$) and a repulsive polaron (blueshifted branch, not labeled). The PL of the moiré layer exhibits abrupt energy shift and intensity suppression at the correlated insulating states. A new correlated insulating state is observed at $\nu = 1$ under $E = 15$ and 18 mV/nm; it can no longer be identified at $E = 30$ mV/nm.

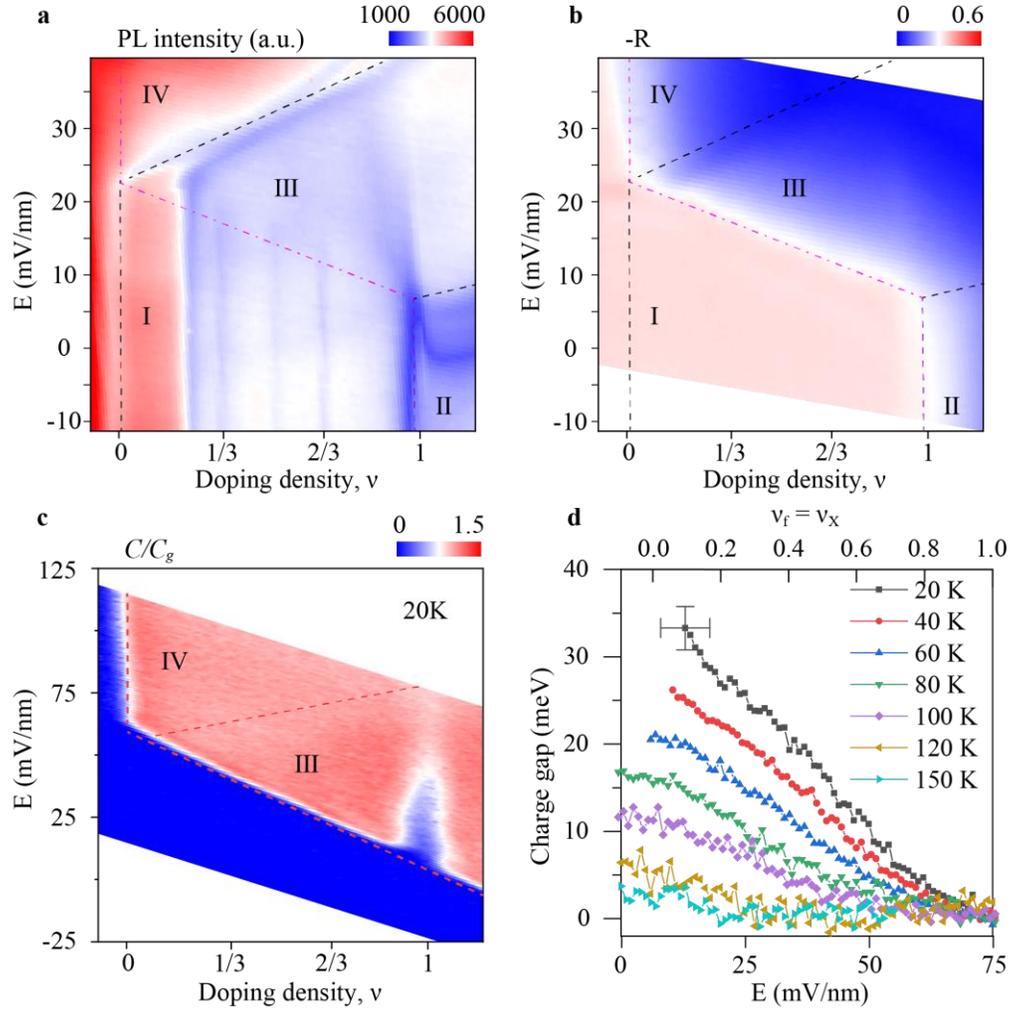

**Figure 3 | Excitonic insulator at ν = 1.** Peak intensity of the moiré layer PL (**a**) and peak reflection contrast amplitude of the monolayer 1s exciton (**b**) as a function of doping density and electric field from device 1. The four regions (I-IV) are defined by the electrostatic doping density in the monolayer ($\nu_f$) and the moiré layer ($\nu_m$). Region I: $\nu_f$ = 0, 0 < $\nu_m$ < 1; Region II: $\nu_f$ > 0, $\nu_m$ = 1; Region III: $\nu_f$ > 0, 0 < $\nu_m$ < 1; Region IV: $\nu_f$ > 0, $\nu_m$ = 0. The boundaries determined by the data in **a** and **b** are shown in dashed and dashdotted lines, respectively. The boundary between region I and II can be determined from either **a** or **b**. A new correlated insulating state is observed at ν = 1 in region III. **c.** Differential bottom-gate capacitance $C$, in units of the geometrical capacitance $C_g$, as a function of (ν, E) at 20 K from device 2. Only region III and IV are accessible because of the device structure. A charge-incompressible state is observed at total filling ν = 1. **d.** Charge gap extracted from the capacitance measurement as a function of electric field (bottom axis) and doping density in the monolayer (or exciton density, top axis) at varying temperatures. The typical error bars are the propagated uncertainty in the gate-dependent capacitance (vertical) and in the optical determination of the boundary (horizontal).

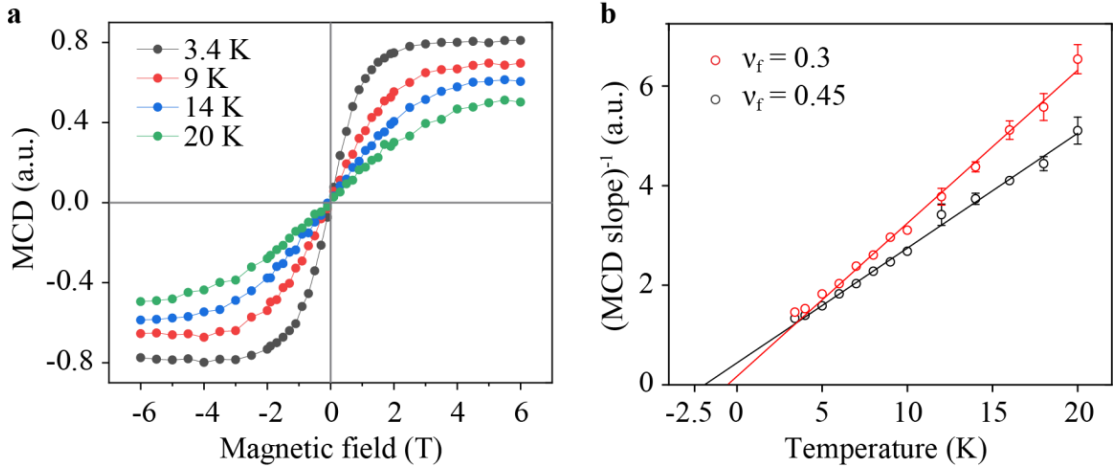

**Figure 4 | Emergence of local magnetic moments in monolayer WSe₂. a.** Out-of-plane magnetic-field dependence of MCD at $\nu = 1$ and $\nu_f = \nu_X \approx 0.3$ (at which the heterostructure is an excitonic insulator) for varying temperatures. The spectral dependence of the MCD is determined as the difference between the left- and right-handed reflected light intensity divided by the total reflected light intensity. It is integrated over a narrow spectral range near the repulsive polaron resonance (1.73 eV) as detailed in Methods. **b.** Temperature dependence of the inverse MCD slope around zero magnetic field (proportional to the inverse magnetic susceptibility) as a function of temperature at $\nu_f \approx 0.3$ and 0.45. The error bars are estimated from the linear fit to the low-field data in **a**. The magnetic susceptibility follows the Curie-Weiss dependence (solid lines) at high temperatures with small negative Weiss constants (the x-intercepts). The magnitude of the Weiss constant is compatible with weak antiferromagnetic exchange interactions between the distant local moments.

**Extended data figures**

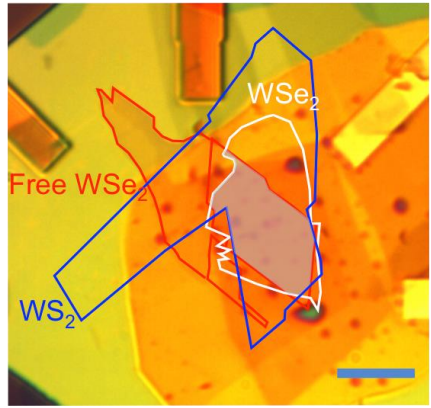

**Extended Data Figure 1 | Optical micrograph of device 1.** The constituent layers are outlined by lines of different colors and labeled. The effective device area is shaded in grey. The scale bar is 10 μm.

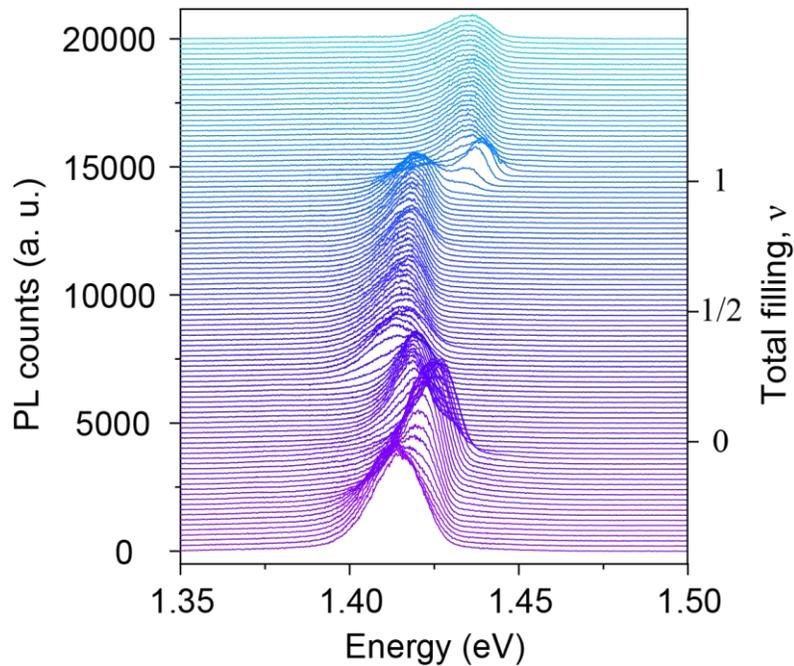

**Extended Data Figure 2 | PL spectrum versus hole doping density.** PL spectrum of the heterostructure, corresponding to the data in Fig. 1f, at different hole doping densities at $E$ = - 34 mV/nm. The spectra are vertically displaced for clarity. The right axis denotes the corresponding hole doping density. An abrupt energy shift is observed at $\nu$ = 1.

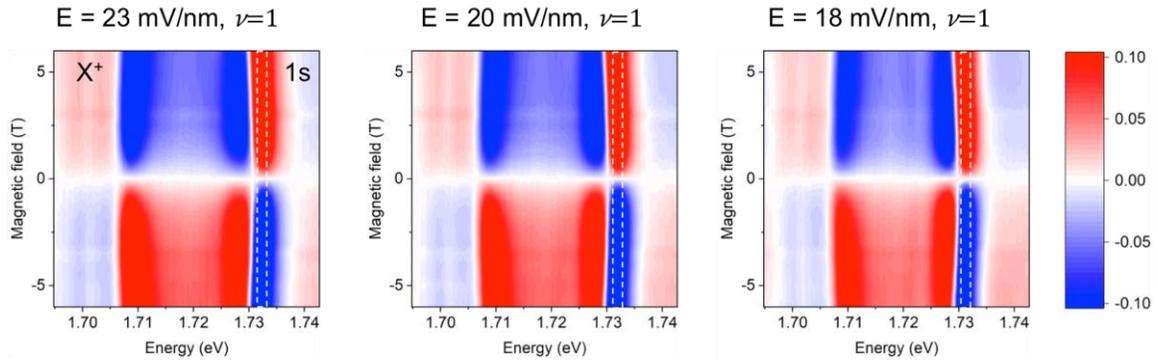

**Extended Data Figure 3 | MCD spectrum as a function of magnetic field at $\nu = 1$ and three representative electric fields $E$.** The spectra focus on the attractive and repulsive polaron resonances of the 1s exciton of the WSe$_2$ monolayer. The MCD is enhanced at the resonances. The box shows the spectral window, over which the MCD is integrated. The magnetic-field dependence of the integrated MCD is shown in Fig. 4 of the main text.

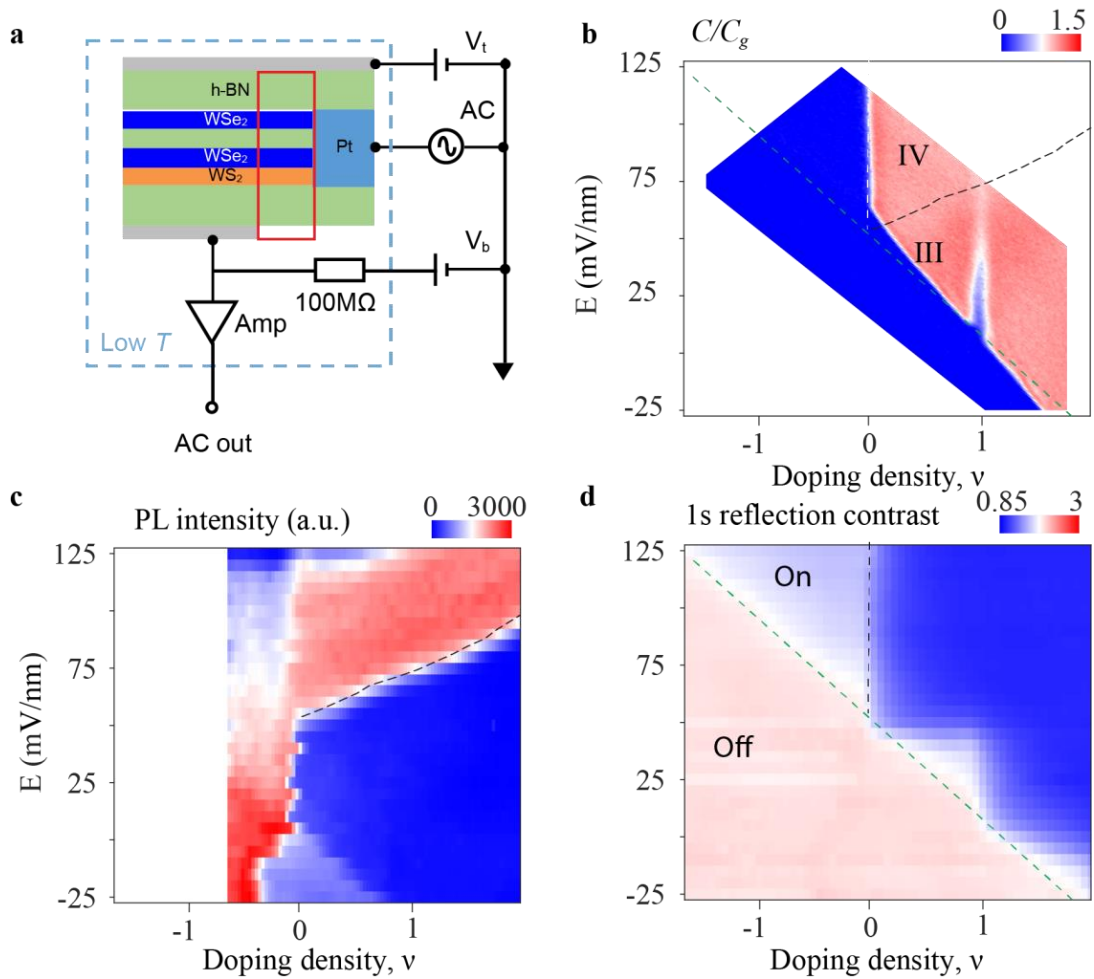

**Extended Data Figure 4 | Capacitance measurements. a.** Schematic of device structure and electrical connections for the AC capacitance measurement. **b-d.** Differential capacitance (**b**), PL peak intensity of the moiré layer (**c**), and 1s exciton peak reflection contrast of the monolayer (**d**) as a function of doping density and electric field ($\nu$, $E$). Electrostatic doping region III and IV in **b** are determined by the optical responses in **c** and **d**. A charge-incompressible state is observed at $\nu = 1$ in region III, which weakens with increasing electric field. The dashed line in **c** marks the onset of doping into the moiré layer; it separates region III and IV. The black and green dashed line in **d** mark the onset of charging into the monolayer and in the device, respectively.

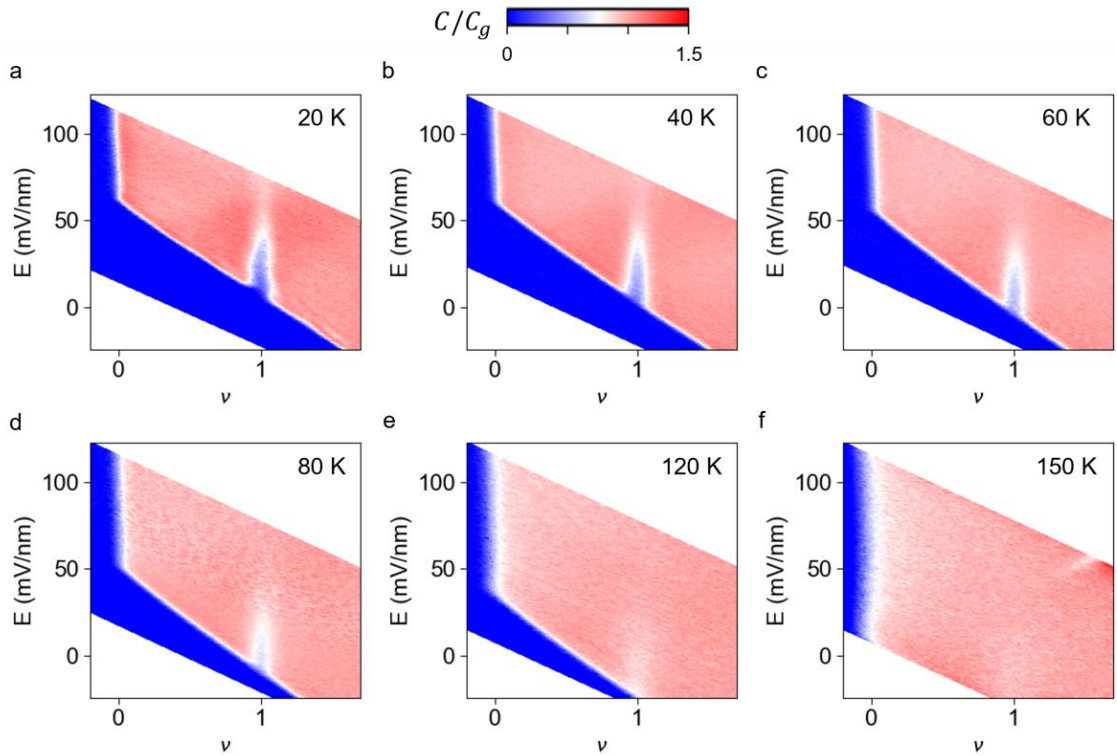

**Extended Data Figure 5 | Temperature dependence of differential capacitance.** Differential top-gate capacitance as a function of $(\nu, E)$ at 20 K (**a**), 40 K (**b**), 60 K (**c**), 80 K (**d**), 120 K (**e**), and 150 K (**f**). The charge-incompressible state at $\nu = 1$ disappears around 120 K.

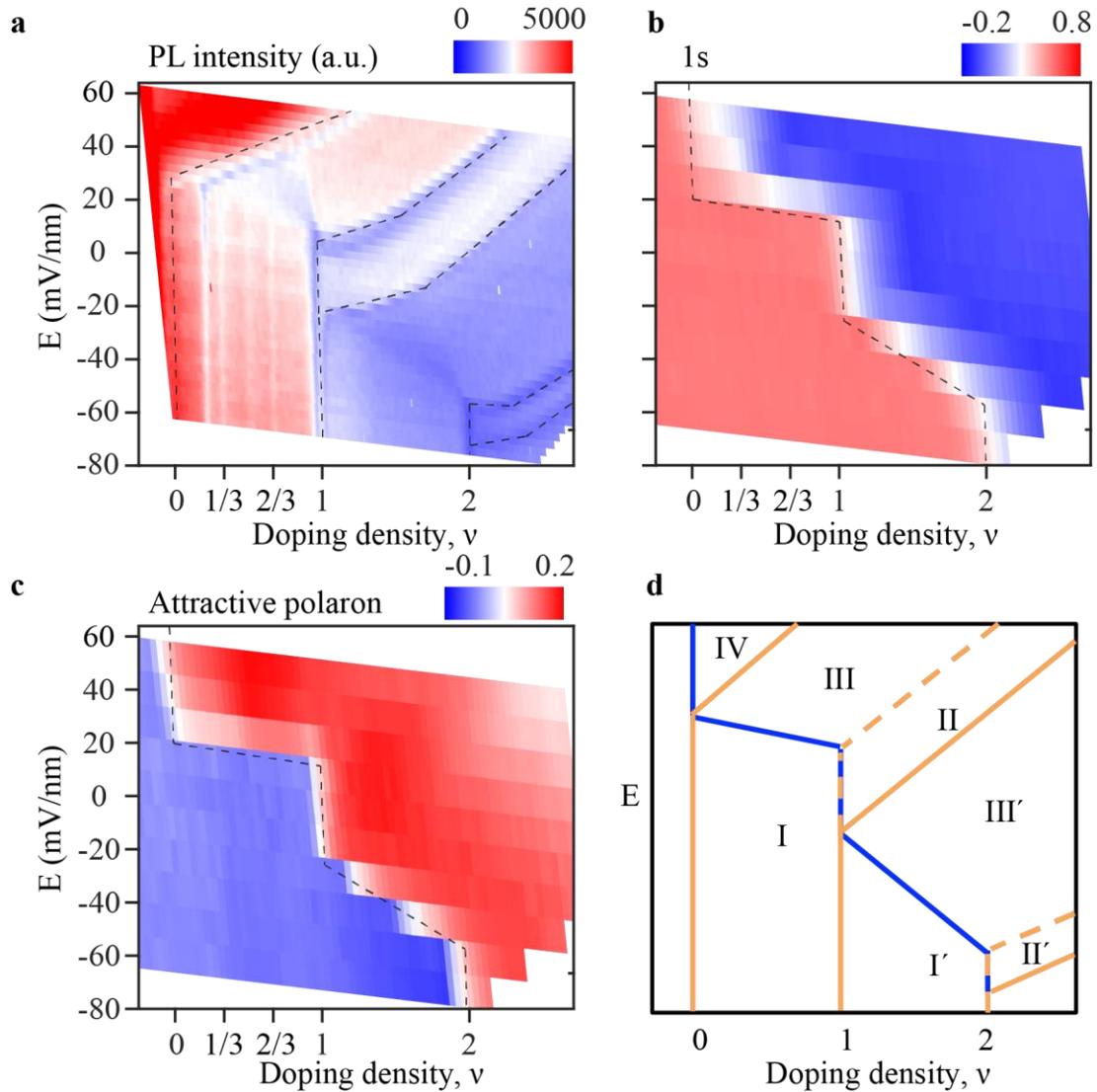

**Extended Data Figure 6 | Electrostatics phase diagram. a-c.** PL peak intensity of the moiré layer (**a**) and the reflection contrast of the 1s exciton (**b**) and the attractive polaron (**c**) of the monolayer as a function of doping density and out-of-plane electric field ($\nu, E$). The PL arises from recombination of the interlayer excitons in the moiré layer. The dashed lines are the boundaries that separate the different electrostatic doping regions (labeled in **d**) and are determined from the optical responses. The data in **a-c** are well correlated with one another. **d.** A schematic of the electrostatics phase diagram of the TMD heterostructure as described in Methods.

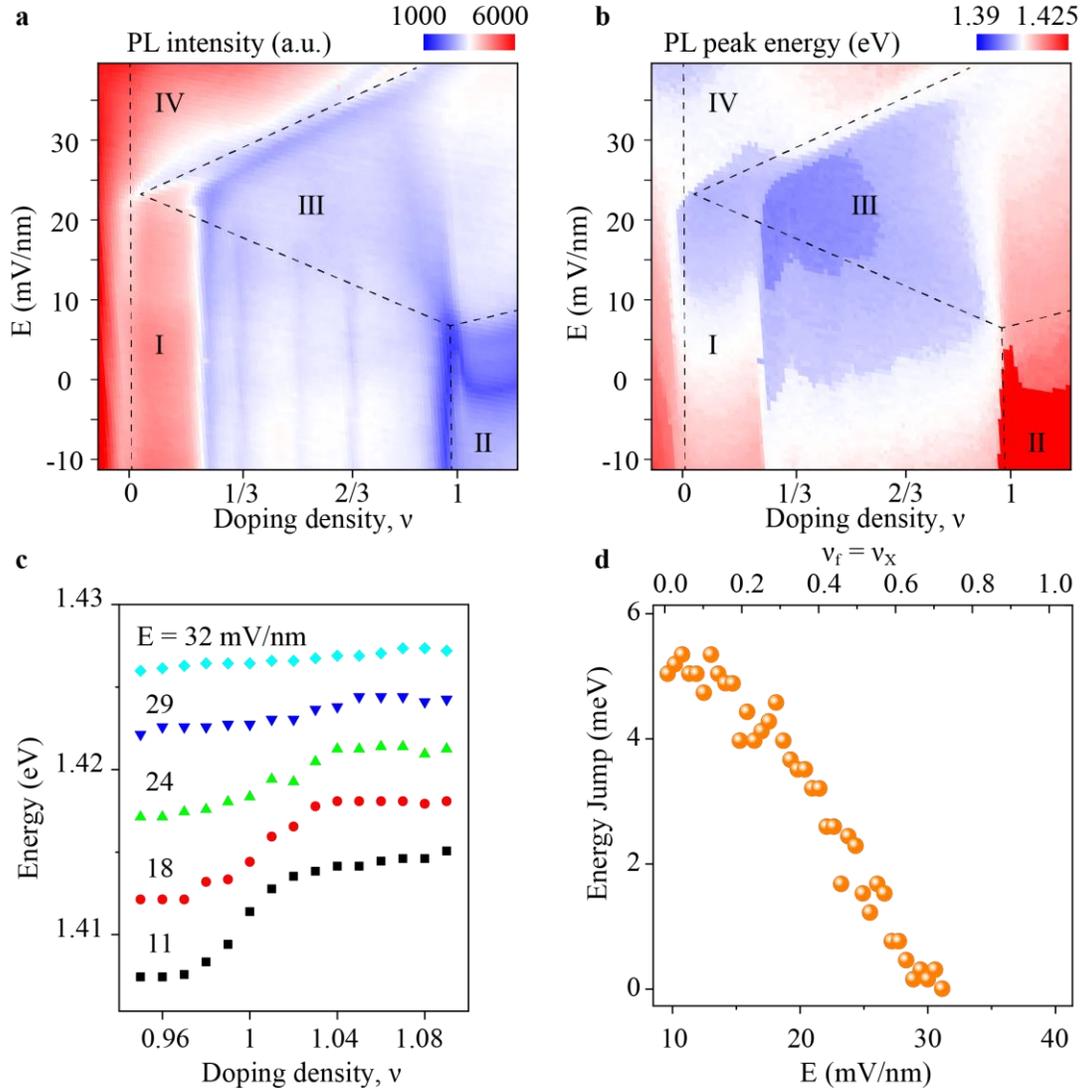

**Extended Data Figure 7 | PL signatures of excitonic insulator.** The PL peak intensity (**a**) and peak energy (**b**) of the WSe₂/WS₂ interlayer moiré excitons as a function of doping density and out-of-plane electric field ($\nu$, $E$). The different electrostatic doping regions are labeled. The feature at $\nu = 1$ in region III corresponds to the correlated excitonic insulating state. **c.** Doping dependence of the PL peak energy at varying electric fields. The abrupt energy shift at $\nu = 1$ reflects a chemical potential jump for the correlated insulating state. The energy shift disappears for the field above a critical value around 30 mV/nm, at which the exciton Mott density (~ 0.7 moiré density) is reached. **d.** Dependence of the PL peak shift at $\nu = 1$ as a function of electric field (bottom axis) and exciton density in the units of the moiré density (top axis). The Mott density is consistent with the value from the capacitance measurement. The smaller electric field range here is due to the thicker hBN spacer compared to the capacitance device.